\documentclass[conference]{IEEEtran}
\pdfoutput=1
\IEEEoverridecommandlockouts
\usepackage{cite}
\usepackage{amsmath,amssymb,amsfonts}
\usepackage{algorithmic}
\usepackage[pdftex]{graphicx}
\graphicspath{{./Figures/}}
\usepackage{wrapfig}
\usepackage{textcomp}
\usepackage{xcolor}
\usepackage{comment}

\usepackage{pdfpages}
\usepackage{hyperref}
\usepackage[absolute]{textpos}
\usepackage{pdfpages}

\usepackage{booktabs}
\usepackage{siunitx}
\newcommand*{\priority}[1]{\raisebox{-1pt}[0pt][0pt]{\begin{tikzpicture}[scale=0.125]%
    \draw (0,0) circle (1);
    \fill[fill opacity=0.5,fill=black] (0,0) -- (90:1) arc (90:90-#1*3.6:1) -- cycle;
    \end{tikzpicture}}}
\usepackage[utf8]{inputenc}
\usepackage{etoolbox}
\usepackage{tkz-euclide}
\usepackage{pdfpages}

\usepackage{url}
\usepackage{multirow}
\usepackage{makecell}
\def\BibTeX{{\rm B\kern-.05em{\sc i\kern-.025em b}\kern-.08em
    T\kern-.1667em\lower.7ex\hbox{E}\kern-.125emX}}

\usepackage{flushend}
\IEEEoverridecommandlockouts

\begin{document}
\begin{textblock}{24}(1,0.2)
\noindent\tiny  This paper is a preprint; it has been accepted for publication in 2020 6th IEEE Conference on Network Softwarization (NetSoft), 29 June-3 July 2020, Ghent, Belgium   \\
\textbf{IEEE copyright notice} \textcopyright 2020 IEEE. Personal use of this material is permitted. Permission from IEEE must be obtained for all other uses, in any current or future media, including reprinting/republishing this material for advertising or promotional purposes,\\ creating new collective works, for resale or redistribution to servers or lists, or reuse of any copyrighted component of this work in other works.
\end{textblock}
\bstctlcite{IEEEexample:BSTcontrol}

\title{On the Security of Permissioned Blockchain Solutions for IoT Applications%
\thanks{%
\protect\begin{wrapfigure}[3]{l}{.9cm}%
\protect\raisebox{-12.5pt}[0pt][0pt]{\protect\includegraphics[height=.8cm]{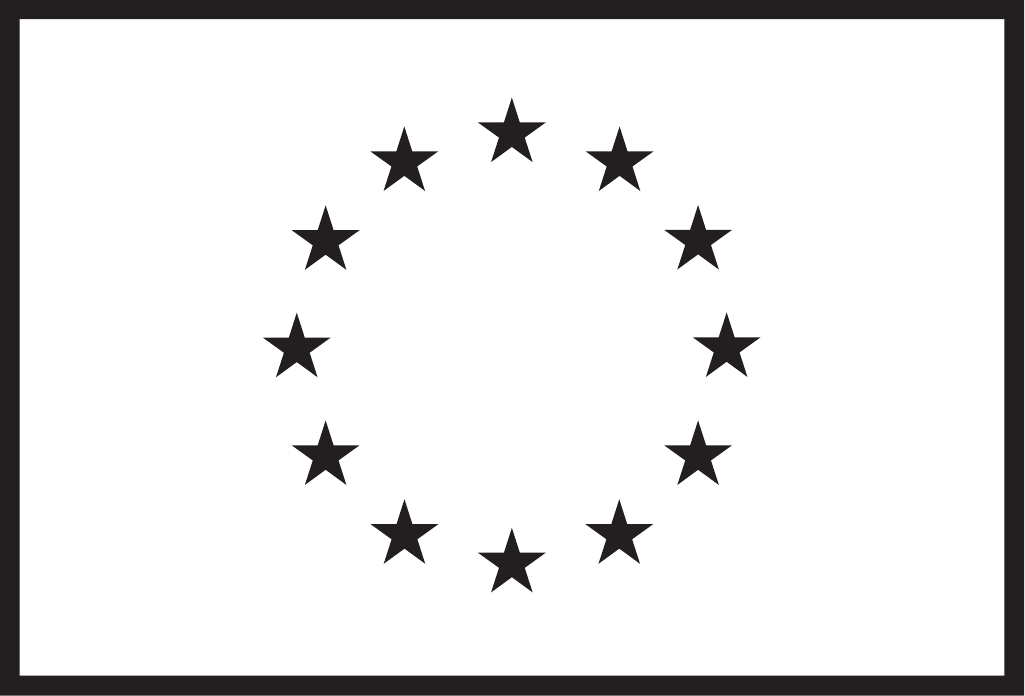}}%
\protect\end{wrapfigure}%
This project has received funding from the European Union's Horizon 2020 research and innovation programme under grant agreement no. 786698. The work reflects only the authors' view and the Agency is not responsible for any use that may be made of the information it contains.}}

\author{%
\IEEEauthorblockN{%
Sotirios Brotsis\IEEEauthorrefmark{1},
Nicholas Kolokotronis\IEEEauthorrefmark{1},
Konstantinos Limniotis\IEEEauthorrefmark{2},
Stavros Shiaeles\IEEEauthorrefmark{3}
\vspace*{4pt}}
\IEEEauthorblockA{\IEEEauthorrefmark{1}University of Peloponnese, Greece. Email: \{brotsis,\,nkolok\}@uop.gr}
\IEEEauthorblockA{\IEEEauthorrefmark{2}Hellenic Data Protection Authority, Greece. Email: klimniotis@dpa.gr}
\IEEEauthorblockA{\IEEEauthorrefmark{3}University of Portsmouth, UK. Email: stavros.shiaeles@port.ac.uk}
}

\maketitle


\begin{abstract}
The blockchain has found numerous applications in many areas with the expectation to significantly enhance their security. The \textit{Internet of things} (IoT) constitutes a prominent application domain of blockchain, with a number of architectures having been proposed for improving not only security but also properties like transparency and auditability. However, many blockchain solutions suffer from inherent constraints associated with the consensus protocol used. These constraints are mostly inherited by the permissionless setting, e.g. computational power in proof-of-work, and become serious obstacles in a resource-constrained IoT environment. Moreover, consensus protocols with low throughput or high latency are not suitable for IoT networks where massive volumes of data are generated. Thus, in this paper we focus on permissioned blockchain platforms and investigate the consensus protocols used, aiming at evaluating their performance and fault tolerance as the main selection criteria for (in principle highly insecure) IoT ecosystem. The results of the paper provide new insights on the essential differences of various consensus protocols and their capacity to meet IoT needs.

\end{abstract}



\section{Introduction}
\label{sec:intro}

The blockchain is an innovative technology, which has proven to provide considerable advancements in several areas by inherently providing interoperability, security, privacy  and sustainability; it can be used as an independent root-of-trust in an distributed (possibly adversarial) setting to allow a set of entities mutually trust each other. For this reason, the blockchain is considered as a decentralized technology for the sharing of information and performing transactions in a secure manner. In contrast to permissionless blockchains, in which any node can participate in the network (\texttt{Ethereum}, \texttt{Bitcoin}, etc.), the permissioned blockchains are characterized by the fact that the participating nodes who regulate and amend the state of the ledger can be identified and can be held accountable for their actions. The key idea behind this concept is to address the privacy needs and other requirements of distributed applications, as well as to provide convenient access control mechanisms with some means of identification.

The current state-of-the-art in permissioned blockchains is rather to adopt the design decisions made in permissionless architectures \cite{Rethinking}. Unfortunately, in most cases this approach leads to sub-optimal performance since a well-defined protocol under the rules of a public ledger,  often with financial orientation, might not be well addressing the challenges of private networks and IoT environments. The transactions' fees, the (high) processing time and (low) networks' throughput should not be inherited by the consensus protocols of permissioned blockchains. Apart from the above constraints, the IoT devices should not be part of the consensus process themselves, due to the fact that several consensus protocols demand considerable computational resources that smart devices do not possess. These issues have been identified in numerous permissioned blockchains \cite{Rethinking}, where the design decisions inherited by the permissionless settings pose great threats to the security, performance, and eventually the blockchain solution's suitability in demanding IoT applications.

Apart from several attacks in blockchains and distributed ledgers, like double-spending and selfish mining, permissioned blockchains often rely on low-level trust models for validation that are originating from the consensus protocol and are hard to be adapted to the operation of smart contracts \cite{Androulaki}. In addition, nodes involved in the consensus protocol might stop working, start behaving in a malicious manner, or acting selfishly against common goals and thus undermine the security of the protocol. Therefore, the consensus nodes should execute a fault-tolerant protocol to safeguard the integrity of the transactions as well as the total order in which they are included in the blockchain in order to deliver a continuous service.

Motivated by the above concerns, and the need for enforcing accountability in many applications, this paper focuses on permissioned blockchains and analyses the properties of a large number of state-of-the-art consensus protocols. Accountability is achieved either directly by immediately proving that a node behaved in a dishonest way, or indirectly by inferring possible malicious behavior using side information at some later stage; trust management schemes can be used to expose long-term malicious trends that are difficult to distinguish at any given time if considered independently. Our work complements the reports by Cachin \cite{Cachin}, Xiao \cite{Xiao}, Zheng \cite{Zheng}, Wu \cite{Wu}, Salimitari \cite{Salimitari}, Lao \cite{Lao}, and Ferdous \cite{Ferdous}, by considerably extending the number of blockchain platforms and consensus protocols that are studied, by providing quantitative information about their performance (i.e. throughput and latency) instead of qualitative information that is given in all works (thus, allowing to easily consider issues related to scalability), and by evaluating their suitability for IoT applications also taking into account their resilience to adversarial faults. The comparative evaluation of the various consensus protocols suggests that only few of them meet IoT needs, in addition to achieving high security.

The paper is next organized as follows. Section \ref{Fault Tolerance in Blockchains} introduces the basics about consensus protocols' security against faults. In Section 
\ref{Crash Fault Tolerant Consensus} and \ref{Byzantine Fault Tolerant Consensus} we give a detailed analysis of the most prominent crash fault tolerant and Byzantine fault tolerant consensus protocols for permissioned blockchain platforms. The comparative evaluation of the various protocols is provided in Section \ref{Comperative evaluation}, whereas Section \ref{sec:conclusion} concludes the paper and discusses future research steps.


\section{Fault tolerance properties}
\label{Fault Tolerance in Blockchains}

Distributed systems are often disciplined by a set of clients and services, where each service utilizes one or more servers to extract information or execute operations that are requested from the clients. Using a central server is the easiest way to fulfill the necessary needs  and the implementation of a service, but it posses a major concern about the security of an IoT network, due to the fact that it becomes a \textit{single point of failure} (SPoF). Thus, to avoid centralized faults, multiple servers should be deployed to implement a Fault Tolerant service with \textit{state machine replication} (SMR) \cite{Schneider}.

The pioneering work of Lamport \cite{Lamport}, who introduced the Byzantine Agreement, triggered the  research of developing algorithms in order to exploit and construct resilient distributed systems.  Nonetheless, several blockchain systems  deviate from the classical SMR in crucial ways. Many distributed applications run simultaneously and can be deployed at any time, even if the embedded application code is un-trusted or occasionally malicious. The key idea to provide security in  blockchains is to reach agreement on a single request from a client, which is the core functionality of  a consensus protocol. In the context of blockchains, a consensus protocol or as it is commonly known as an ``atomic broadcast'', provides a total order of the disseminated messages and propagates them to the network peers.


\subsection{Crash fault tolerance}
\label{Crash Fault Tolerance}
An atomic broadcast certifies that all the legitimate nodes output or deliver the identical array of  messages by means of the deliver event. Accurately, considering a set of $n$ nodes in the network, it certifies that the properties validity, integrity and  total order are fulfilled \cite{Hadzilacos}. 

The way to achieve consensus (i.e. to realize atomic broadcast) in distributed systems that are vulnerable to $t< n/2 $ node crashes is to adopt  consensus protocols known as the  \textit{viewstamped replication} (VSR) \cite{Oki} and  \texttt{Paxos} \cite{Lamport2} family of protocols.  To provide security, this family  is characterized by the same rules. In each round a leader is elected or voted to create a new block and if the ongoing leader crashes or even if the nodes in the network suspect that the leader has crashed, the leader is reinstated by proceeding to the next round. This family of protocols is known today as  \textit{crash fault tolerant} (CFT) consensus protocols and it guaranties that a set of failing nodes $t< n/2 $ does not impact the system.


\subsection{Byzantine fault tolerance}
\label{Byzantine Fault Tolerance}

Consensus protocols with the purpose of tolerating byzantine nodes, which are subverted by a malicious actor and avert the common goal of reaching agreement, have recently emerged. In the  \textit{Byzantine fault tolerant} (BFT) consensus protocols family, the most common protocol is the  \textit{practical Byzantine fault tolerant} (\texttt{PBFT}) \cite{Castro}, which can be displayed as a blossom of the \texttt{VSR}/\texttt{Paxos} \cite{Oki}, \cite{Lamport2}. In a network comprised of a set of $n$ nodes, the \texttt{PBFT} consensus protocol \cite{Castro} can tolerate $f<n/3$ subverted nodes using a progression of rounds with a unique leader within each round. Under the assumption  that the BFT protocols seem to be more secure than CFT, vast research work has focused on the  improvement of the \texttt{PBFT} with \texttt{BFT-SMaRt} \cite{Bessani} to be considered one of the most advanced and scalable BFT consensus protocols.

\section{Crash fault tolerant consensus}
\label{Crash Fault Tolerant Consensus}

Fault-Tolerant algorithms have acquired momentous observation over the years \cite{lynch1996distributed}. Some of them provide significant results  of solving consensus, in an optimum way, by identifying bounds on the security beneath different models. In this section, we survey recent CFT consensus protocols, focusing on results for permissioned, IoT-based blockchain platforms. However, these protocols cannot endure  malicious activities but they can only   tolerate  $50\%$  of the network crashes,  making their adoption for IoT networks not a very appropriate choice. 


\subsection{Kafka / Zookeeper: Hyperledger Fabric, Corda} 
\label{Apache Kafka} 

\texttt{Kafka} \cite{Kafka} is a distributed publish-subscribe streaming platform, embraced by  \texttt{Fabric}  \cite{Androulaki} and experimentally by  \texttt{Corda} as a cluster of ordering/notary service. At high level, the conceptual configuration of  \texttt{Kafka} is identical with the leader -- follower setting. The transactions, which  \texttt{Kafka} calls \textit{messages}, are replicated from the leader to its followers and if the leader crashes, then one of the followers takes command. This action ensures the crash-fault-tolerance of the network. The administration of  \texttt{Kafka} \cite{Kafka}, including the systematization of tasks,  the  election process, the cluster association, among others,  is organized by \texttt{Zookeeper} \cite{ZooKeeper}. Despite the fact that \texttt{Kafka} is CFT, \texttt{Zookeeper} by drifting \texttt{Kafka} becomes somehow centralized and thus the protocol is not advised  to be deployed in a large network. In simple words, the ordering service of  \texttt{Fabric} is executed by distinct organizations and  \texttt{Kafka} demands one of them to run the cluster. This means that all the ordering  nodes are attracted to the same  cluster, which is under the jurisdiction of a single organization. Thus, this concept does not assure much in terms of decentralization, even though that the protocol achieves a significant throughput with very low latency \cite{Androulaki}. 


\subsection{Raft: Hyperledger Fabric, Sawtooth, Quorum, Corda}
\label{Raft}

\texttt{Raft} \cite{raft} is a CFT consensus protocol, alternative to  \texttt{Paxos} \cite{Lamport2},  implemented in  \texttt{Hyperledger Fabric} \cite{Androulaki},  \texttt{Hyperledger Sawtooth} \cite{sawtoothdoc}, \texttt{Quorum} \cite{Quorum} and  \texttt{Corda} \cite{Corda1}. As a voting system,  \texttt{Raft}  realizes the leader-follower model, in which each node is characterized as \textit{leader}, \textit{follower} or \textit{candidate}. The consensus process is parted into the following sub-problems: a) the leader election, with  the initiation of the protocol or in case the ongoing leader crashes; b) the  log replication, in which the leader affirms log entries, replicates them  and compels all the followers to acknowledge them and c) the safety of the protocol,  concerning the regulations imposed on the leader-election process to ensure this security property \cite{raft}. The time in  \texttt{Raft} is separated into \textit{terms}. The terms are measured with increasing integers  and initiated with an election. If a candidate wins the election, then it takes the leader's place for this term. The leader disseminates systematic heartbeat messages to its followers and if, for a preordained time period a candidate obtains no heartbeat messages,  then it is presumed that the leader has crashed, which means that a new election process has to be initiated. 

 Although \texttt{Raft} is a fast CFT consensus protocol, it provides low and medium   performance if it is implemented in  \texttt{Corda}  \cite{Ampel} and \texttt{Quorum} \cite{Quorumperformance} (respectively).  On the other hand, in the new version of \texttt{Fabric} (Fabric v2.0) \cite{fabric2.0}, in  which \texttt{Raft} is the recommended protocol, it is possible to achieve twice the transaction  throughput of \texttt{Kafka}'s and thousands of transactions in real world scenarios \cite{fabricraft}, \cite{fabric2.0} which makes  \texttt{Fabric} with \texttt{Raft} a very good candidate for IoT applications.


\section{Byzantine fault tolerant consensus}
\label{Byzantine Fault Tolerant Consensus}

A primary case of  failures that is essential to blockchains, is the malicious behavior caused by an adversary. The failure of nodes is stated, not only on unintentional crashes, but also on contemplated attacks in the system, that attempt to defeat the security enhancements  by causing at least forks in the ledger. In the context of blockchains, the consensus mechanisms are required to address  malicious, selfish and generally any set of nodes that attempt to alter the protocol, by ensuring the safety and the global state coherence. For this reason, an adoption of a BFT protocol with high performance metrics is more suitable for an IoT environment. 


\subsection{PBFT: Hyperledger Sawtooth}
\label{PBFT} 

\texttt{PBFT} is the first BFT protocol based on SMR and was developed by Castro and Liskov \cite{Castro} in 1999. The protocol has become a synonym of BFT and has attained immense concern in distributed consensus. Each participating node is a \textit{validating replica} (VR); one of them is named to be the leader.  The \textit{leader validating replica} (LVR) receives a transaction request from a client, who demands an execution of an operation. Then, the LVR validates the transaction and disseminates this request to the other VRs. In a short period of time, called ``batch-timeout'', or after a plethora of ordered pending transactions, called ``batch-size'', a block is created by the LVR. Then, the block is disseminated to the other VRs to reach consensus, initializing at the same time the \textsf{pre-prepare}, \textsf{prepare} and \textsf{commit} phase of consensus. If $2f+1$ VRs come to an agreement upon a decision, then each VR appends this block as the next block in its private ledger. Although the protocol provides the desirable performance of over $78000$ tps as shown in \cite{Bessani}, it lacks scalability supporting only few tens of participants in the network; demanding high communication complexity  due to the high number of messages that have to be transferred \cite{POA}, \cite{Liu}. These reasons  suggest that the protocol should be deployed in a small IoT environment, for example in a \textit{small office/home office} (SOHO) environment and not in a large network where the high number of nodes can act as bottleneck. 


\subsection{IBFT: Quorum, Hyperledger Besu, Autonity}
\label{IBFT}

\textit{Istanbul Byzantine fault tolerant} (\texttt{IBFT}) \cite{IBFT} is a quite appealing  \textit{proof-of-authority} (\texttt{PoA}) mechanism, which is emanated from the  \texttt{PBFT} and inherits its security properties from it, by employing the three-phase consensus. In a similar way to  \texttt{PBFT}, the VRs elect a LVR to create  a proposed block and disseminate it to the network, along with a \textsf{pre-prepare} message. The VRs, upon accepting this \textsf{pre-prepare} message and to be certain that they are on the same sequence, they enter the \textsf{pre-prepared} phase and  disseminate a \textsf{prepare} message. The LVR, in its turn, while gathering a number of $2f+1$ \textsf{prepare} messages from the VRs,  enters the \textsf{prepared} phase and disseminates a  \textsf{commit} message. This means  that the LVR accepts the proposed block, while declaring that this block is going to be inserted to its ledger. Then, the VRs, upon receiving $2f+1$ \textsf{commit} messages, they enter the \textsf{committed} phase  and accept the proposed block. Although, \texttt{IBFT} is similar to  \texttt{PBFT},  the later needs some tweaks to be used in blockchains. In  \texttt{IBFT}, the fact that there is no particular client to demand results by sending requests, makes all the VRs to be recognized as such. The performance evaluation of Baliga et. al., \cite{Quorumperformance} for using \texttt{IBFT} and \texttt{Raft} in the \texttt{Quorum} platform is not peer-reviewed, but it shows that when private smart contracts are used, in which it is required further   encryption - decryption functions and further communication  overhead  among  peers, it can be achieved a throughput of $600-650$ tps with a latency of around $4.5$sec.  

\subsection{BFT-SMaRT: Hyperledger Fabric, Corda, Symbiont}
\label{BFT-SMaRT}

 Bessani's protocol is a  java-based library implemented in various platforms such as  \texttt{Hyperledger Fabric} \cite{Androulaki},  \texttt{R3 Corda}  (experimentally) and  \texttt{Symbiont Assembly} \cite{Symbiont}. In the absence of malicious VRs,  \texttt{BFT-SMaRT} \cite{Bessani} achieves consensus  using the message model of \texttt{PBFT} \cite{Castro}, but in their presence the protocol forces the network to elect a new LVR and execute the  messaging pattern, which is described in \cite{Bessani}. The differences between the two BFT protocols is firstly that  \texttt{BFT-SMaRT} has enhanced  reliability and  multi-core  processing concerning the appraisal of signatures.  \texttt{BFT-SMaRT} realizes a modular consensus protocol, which is not embodied inside of the SMR while the state transfer and reconfiguration modules are clearly detached from the agreement method \cite{Bessani}. The supported  reconfiguration makes the protocol to differ from previous BFT systems, in which the size of the network could not expand or reduced, and allows VRs to join or leave the network on-the-fly.  In the context of  \texttt{Fabric}, the consensus protocol also implements Wheat \cite{Sousa},  a vote assignment model (as the VRs recognize each other), to achieve the desirable for IoT networks performance without imperiling the safety and the stability of the consensus. Although, the protocol is not in a stable mode if it is implemented in  \texttt{Corda}, in \texttt{Fabric}'s and \texttt{Symbiont}'s case the desirable performance for IoT deployment has been achieved far more than necessary.


\subsection{RBFT: Ontology, Hyperledger Indy}
\label{RBFT}

\texttt{RBFT} which stands for  \textit{redundant Byzantine fault tolerance} is a consensus algorithm proposed by Aublin \textit{et al.} \cite{Aublin} and implemented in  \texttt{Ontology} \cite{Ontology} and in \texttt{Indy} \cite{Indy}. The recent BFT protocols elect in each instance a primary replica, herein called LVR, which demonstrates how  all the incoming requests will be ordered.  \texttt{RBFT} introduces a new approach, running multiple instances concurrently, in order to identify and quickly replace malicious LVRs, with the reduction  of throughput's degradation caused by their presence as its aftereffect. While multiple instances are executed and order requests, their performance is strictly monitored and only the requests that have been ordered by the master instance are executed. The detection of a malicious master LVR, which wants to smartly and willingly degrade the performance of the protocol, results from comparing its performance with all the other LVRs' performances. This comparison takes place due to the fact that in BFT protocols it is hard for VRs to presume the throughput that a non-malicious LVR would have. In the case, where a LVR is faster that the master LVR, then a new LVR is elected in each instance and the master LVR is marked as malicious. In \texttt{RBFT} \cite{Aublin}, when malicious faults occur, the throughput is degraded only by $3\%$, while the degradation of other existing BFT protocols, under the same faults, is at the best of circumstances equivalent to $78\%$. This property makes the protocol a quite appealing mechanism to be used in IoT environments; if it is combined with a  platform with high  performance.  


\subsection{VBFT: Ontology}
\label{VBFT}

\texttt{VBFT} \cite{VBFT}, the core consensus protocol of  \texttt{Ontology}, \cite{Ontology},  relies on \textit{verifiable random functions} (\texttt{VRF}) \cite{VFR} to introduce randomness and is combined with  \textit{proof-of-stake} (\texttt{PoS}) and  Byzantine tolerance to provide resistance against malicious acts. The whole network consists of two different types: The consensus nodes, where the stake of each participant has significant impact; and the consensus candidates, in which the nodes do not aid the consensus process, but  validate consensus blocks and update their ledger. Due to the randomness provided by the  \texttt{VRF} function, the  \texttt{VBFT} protocol selects different sets of nodes, which are difficult to predict and  each one of them is assigned a different work to perform. At first, each proposal node creates and proposes a new block.  The proposed blocks are collected from the verification nodes, which verify  and independently vote for them depending on their highest priority. Then, the confirmation nodes confirm the  results that were provided by the verification nodes and finally establish a consensus result. Upon the reception of an  established result, a new round begins.  The protocol is only implemented in the \texttt{Ontology} \cite{Ontology} platform -- a public, high performance and scalable permissioned blockchain - that enables   smart contracts  for different business requirements. 


\subsection{Tendermint: Hyperledger Burrow, Autonity, Ethermint}
\label{Tendermint}

\texttt{Tendermint} \cite{Tendermint}, which is majorly adopted in \texttt{Burrow} \cite{Burrow},  \texttt{Autonity} \cite{Autonity} and  \texttt{Ethermint} \cite{Ethermint}, is designed as a deterministic  protocol under a partially synchronous communication. Although the  \texttt{Tendermint} consensus has similarities to other BFT protocols, the voting power in this protocol differs and  is defined based on each node's stake. Particularly in the \textsf{propose} phase, a VR is deterministically selected, according to the rate of its stake, to propose a block for a specific height. Then, for a block to be committed, the \textsf{pre-vote}, \textsf{pre-commit} and the \textsf{commit} phases  have to follow in a way similar to most BFT protocols. 

Furthermore, a concept called \textit{locks} or \textit{polka} is included in \texttt{Tendermint}'s terminology. More precisely, in the \textsf{pre-commit} phase, if a proposed block gathers more than the $2/3$  of the VRs' \textsf{pre-votes} for a block, then a lock on the proposed VR occurs. For various reasons, a new block may fail to be committed in a specific height. In this case, the protocol moves on to a new round with a new VR to propose a new block for that height. In such a round, the new proposer may be locked in a block from a preceding round. Then the proposed-locked block is the same as before and in its proposal a \textit{proof-of-lock} (PoL) is added. The PoL \cite{Tendermint}  is a collection of \textsf{pre-votes} from the VRs, concerning the situation of a proposed block, whether that is locked or unlocked.

In the \textsf{pre-vote} phase, each locked VR signs and disseminates a \textsf{pre-vote} concerning its locked block. Frequently, due to network asynchrony, the VRs may not obtain any proposal or even worse, obtain an invalid block. In such a case, they sign and then  disseminate a nil \textsf{pre-vote}. Generally, a nil vote, depending on the phase that is taking place, is  a vote to either  move to the next round or to unlock a block and a situation in a later round, when another block is locked for the same height. In the \textsf{pre-commit} phase, if more than $2/3$ from the pre-votes are nil votes, then the VR and the block are considered unlocked and a \textsf{propose} phase of a new round is initiated.  Otherwise, the VRs enter the \textsf{commit} phase. Although \texttt{Tendermint} is a secure BFT consensus protocol, its implementation in \texttt{Ethermint} and \texttt{Burrow} provides low and medium transaction throughput (respectivelly) making the protocol not the best option for IoT environments.


\subsection{Exonum consensus: Exonum}
\label{Exonum Consensus}

The \texttt{Exonum} consensus \cite{Exonum}, a customized BFT algorithm, is built  by Bitfury and  implemented only to the  \texttt{Exonum} platform. The network comprises two different kinds of nodes, with each kind to have been assigned a different work to perform; on the one hand the Auditors, which do not assist in reaching consensus but can read the blocks' transactions and on the other hand the validators, herein called VRs, which participate in the consensus process. As it commonly happens in a BFT protocol, each round is initiated with an elected LVR who sends a  proposal to be embodied in the blockchain. The three-phase consensus (\textsf{pre-vote}, \textsf{pre-commit} and \textsf{commit}) is almost the same for the \texttt{Exonum} consensus. Similarly to Tendermint, a special concept called ``lock'' is added to the protocol. The lock defines that the VR, whose \textsf{pre-vote} block has gather more than $2/3$ of the approvals from the network, gives up voting for other VRs' proposals and locks on its own, without giving any permission to change it. Then, the VR broadcasts a \textsf{pre-vote} message expressing its locked condition on the specific proposal. With high throughput, low latency and the support of smart contracts, \texttt{Exonum} makes a good candidate to be implemented in IoT networks if the concept of ``locks'' is connected with IoT-based criteria.     


\subsection{DPoS: EOSIO, Bitshares, Tron , Tezos, Lisk}
\label{DPoS}

\textit{Delegated proof-of-stake} (\texttt{DPoS}), a reputation-based consensus protocol, is implemented in  \texttt{Bitshares} \cite{Bitshares}, \texttt{EOSIO} \cite{EOSIO}, \texttt{Tron} \cite{Tron}, \texttt{Tezos} \cite{Tezos}, and \texttt{Lisk} \cite{Lisk}. In  \texttt{DPoS}, the nodes vote with reputation scores to choose a class of delegates that will be assigned to create blocks. In each round, among the set of delegates, a  leader is selected in a way defined by the respective distributed ledger. The leader is incentivized to follow the protocol and get rewarded upon the creation of an honest block and penalized or blacklisted in any other case. Among the delegates, a contest on which one of them is going to be included in the validation set takes place, with each delegate promising various levels of rewards to its voters if it is elected as a leader. Each ledger follows its owns parameters and by using a small number of validators, finality can be achieved promptly. The performance evaluation of \texttt{DPoS} varies across different platforms, with \texttt{EOSIO} and \texttt{Bitshares} to provide a  desirable, suitable and high performance network for IoT environments if the stake is replaced with IoT-based criteria. 


\subsection{Clique / Aura: Quorum, Parity, Autonity}
\label{Clique and Aura}

\texttt{Clique} and \texttt{Aura}  are  \texttt{POA} protocols \cite{POA} developed to supposedly displace the computational consumption in permissioned blockchains and to increase the transaction throughput.  \texttt{Clique} is implemented in  \texttt{Quorum} \cite{Quorum} and  \texttt{Autonity} \cite{Autonity}. The time in  \texttt{Clique} proceeds in epochs, where during each epoch the current mining leader and multiple other authorities are allowed to propose a block after a determined number of epochs. \texttt{Aura}, on the other hand,  is implemented in  \texttt{Parity} \cite{Parity} and does not commit the proposed block at once, but rather adds an extra round, called ``block acceptance'', to examine if the received  block is the same to all the authorities. In cases where the leader has not proposed any block (even empty) or has proposed multiple and contrasting blocks to various authorities, then  the leader's position is under vote. If the leader is emerged as malicious then it is removed. An academic research \cite{POA} showed that under the presence of byzantine authorities   \texttt{Clique} and  \texttt{Aura} may fork and thus result in the violation of the protocol's safety and reduction of the network's throughput. \texttt{Clique} and \texttt{Aura} provide high Byzantine fault tolerance of $2f+1$ nodes, but  as consensus protocols might encounter network delays and block confirmation time over $5$sec, which may not be admissible in delay-delicate IoT environments.

\subsection{DBFT: NEO, Ontology}
\label{Delegated Byzantine Fault Tolerance (DBFT)}

Developed from the  \texttt{NEO} team \cite{Neo} and later adjusted to the \texttt{Ontology} platform \cite{Ontology}, the \textit{delegated Byzantine fault tolerance} (\texttt{DBFT}) protocol follows  the structure of  \texttt{PBFT}  and  \texttt{DPoS}, providing thus the best of both worlds. From  \texttt{PBFT}, the protocol inherits the three-phase consensus without the need for all the nodes to participate in it. With this adjustment the protocol becomes stable and averts malicious forks.   On the other hand, the voting system, in which only the delegates can execute the consensus process, is inherited from the \texttt{DPoS} protocol, providing thus a sampling of the most trusted entities in the network. The derived advantages of the combination of the two are: the instant transaction finality without forks and a fast voting-based  method to elect the delegates. The high transaction throughput  and the security that the protocol provides are enticing features for IoT applications.  However, in \texttt{Ontology} and in \texttt{NEO} platform, the block confirmation time is  $5-20$sec, which is  beyond the appropriate restraints and  makes the protocol not so applicable for  IoT devices.  

\subsection{YAC: Hyperledger Iroha}
\label{YAC}

\textit{Yet another consensus} (\texttt{YAC}) \cite{YAC} is a novel protocol implemented in \texttt{Hyperledger Iroha} \cite{Iroha} with its priority the mobile app development using the simplest possible construction. In the  \texttt{YAC} protocol, the validation process is not part of the consensus, but it is rather  based on the transaction flow of  \texttt{Iroha}. In order to be considered valid, each transaction has to pass at first the stateless validation ---\,in which the transaction's signature and format are checked\,--- and afterwards the stateful validation that  takes place in a slower form, after the  ordering process. The procedures performed by the protocol are limited in ordering and consensus; The first is executed by the ordering service and the later by an ordered list of peers, in which the first peer is considered to be the leader. Although, the  \texttt{Iroha}'s transaction flow  to an extent reminds us of  \texttt{Fabic}'s, here only the verified transactions are included in a block. Upon the reception of the ordered sequence of transactions in a pattern of proposals, the peers perform  the stateful validation check, remove invalid transactions from the sequence of transaction, create a proposed block, vote the proposed block through signing and then disseminate it to the leader. After receiving  $2/3$ of the proposed-voted blocks, the leader sends to all the consensus  peers  a \textsf{commit} message manifesting which block should be accepted to their chains. Although \texttt{Iroha}'s throughput is satisfiable for IoT environments, the block confirmation time is  slightly over from other favorable permissioned blockchains due to the  restraints that small vote delays enforce on the peers in order to come to an agreement on a specific  proposal.

\subsection{PoET: Sawtooth}
\label{PoEt}

To solve the problem of Byzantine agreement and to avoid the wasteful computational power that Nakamoto's consensus utilizes \cite{bitcoin}, \textit{proof-of-elapsed time} (\texttt{PoET}) implements a lottery-based algorithm to achieve fairness, investment and verification in the leader election process. The peers in \texttt{Sawtooth} \cite{sawtoothdoc} are thoughtfully elected to execute requests after waiting an indefinite period of time \cite{PoET}. The  peer with the lowest waiting time creates the new block. To certify that each peer's time has actually elapsed, the protocol demands the whole critical process to be performed in a private memory area called \textit{trusted executed environment} (TEE), i.e.  Intel SGX \cite{SGX}. Specifically, the  TEE's utility, among others, is to  provide a cohesion proof that results from a trusted function, often called ``enclave'', by means of remote attestation in order to establish trust on the consensus network. The protocol is BFT but malicious adversaries can manipulate the network if they compromise participating nodes beyond  the threshold of $\Theta(\frac{\log\log n}{\log n})$ \cite{POETtolerance}. The  protocol  possesses the required security and performance criteria to be deployed in IoT environments, but  without integration of the SGX hardware, \texttt{PoET}'s security degrades into the  family of CFT protocols. 

\subsection{RPCA: Ripple}
\label{Ripple}

The \textit{\texttt{Ripple} consensus algorithm} (\texttt{RPCA}), was introduced in 2014 by Schwartz \textit{et al.}  \cite{Ripple}, as the fundamental protocol for  a secure and fast real-time cryptocurrency-based system to transfer remittances without the support of smart contracts. In the  \texttt{RPCA}, the nodes propose the assembled transactions from their clients to the validators to reach consensus. The time in  \texttt{Ripple} is measured in \textit{epochs}, with each epoch involving several rounds of transactions' refinement.  Initially, with the use of a \textit{unique node list} (UNL), each node identifies all the nodes with which it can instantaneously interact, exchange messages and trust. In a vast network, such as \texttt{Ripple}, trust does not actually mean that each node in a UNL is trusted, but rather that this node will not attempt to circumvent the network with votes on inaccurate transactions. A transaction with a minimum threshold  of yes-votes is forwarded to the next round, while others, that do not exceed the threshold, are either deserted; if malicious, or embodied in a \textit{candidate set} of new and not yet applied transactions on the ledger. The transactions that in the final round exceed with positive votes with a percentage of $80\%$ of the node's UNL, are included to the ledger, while the candidate set is  forwarded for consensus on the next ledger. Taking into account that \texttt{Ripple} is a semi-permissioned system, its cryptocurrency-based nature, its overall performance and the fact that the percentage of the tolerated malicious validators in each node's UNL is  $20\%$; the protocol is not a good option for IoT deployment.  

\label{Comperative evaluation}
\begin{table}[!t]
\setlength{\tabcolsep}{8.5pt}
\renewcommand{\arraystretch}{1.2}
\centering
\caption{Criteria for the suitability of consensus protocols in IoT}
\label{comparisonanalysis}
\begin{tabular}{lcc}
    \toprule
    Criteria & Suitability & Symbol \\
    \midrule
     FT and LP & \textcolor{white}{00}\SI{0}{\percent} & \makecell{\priority{0}} \\
     CFT and MP & \textcolor{white}{0}\SI{25}{\percent} & \makecell{\priority{25}} \\
     BFT and MP & \textcolor{white}{0}\SI{50}{\percent} & \makecell{\priority{50}} \\
     CFT and HP & \textcolor{white}{0}\SI{75}{\percent} & \makecell{\priority{75}} \\
     BFT and HP & \SI{100}{\percent} & \makecell{\priority{100}}\\
    \bottomrule
\end{tabular}
\end{table}


\section{Comparative evaluation}
\label{Comparative evaluation}

The consensus protocols presented in the previous sections have been incorporated into various permissioned blockchain platforms. The transaction throughput and latency are the most critical factors to consider in IoT deployments, as it is required a transaction to be broadcast and appended into a block in a matter of seconds or even less. According to \cite{Salimitari}, throughput and latency are mapped to a three-valued scale, namely \textit{low}, \textit{medium}, and \textit{high}, where the thresholds that define throughput are $100$ and $1000$ \textit{transactions per second} (TPS) respectively; in addition, latency characterization relies on the time units (milliseconds, seconds, or minutes). In this paper, we follow the same three-valued scale but the throughput (resp. latency) thresholds have been adjusted to $500$ and $1500$ TPS (resp. $1$sec and $10$sec) respectively to meet IoT applications' requirements.

\begin{table*}
\setlength{\tabcolsep}{8.5pt}
\renewcommand{\arraystretch}{1.2}
\centering
\caption{Comparative evaluation of blockchain platforms' consensus protocols against the needs of IoT}
\label{comparison}
\begin{tabular}{lllccccc}
\toprule
\makecell{\thead{Consensus\\Protocol}} & \makecell{\thead{Fault\\Tolerance}} & \makecell{\thead{Blockchain\\Platforms}} & \makecell{\thead{Throughput\\(TPS)}} & \makecell{\thead{Latency -- Block\\Confirmation Time}} & \makecell{\thead{Evaluation\\Source}} & \makecell{\thead{Overall\\Performance}} & \makecell{\thead{IoT\\Applicability}} \\

\midrule
 
Aura & BFT: $2f+1$& Parity & $35-45$ & $3-7$sec & \cite{Parity-Quorum-Ethermint}  & Low & \priority{0} \\ \hline

\multirow{2}{*}{BFT-SMaRt} & \multirow{2}{*}{BFT: $3f+1$} & Fabric & $>10\text{K}$ & $0.5$sec & \cite{Sousa} & High & \priority{100} \\

& & Symbiont & $80\text{K}$ & $<1$sec & \cite{Cachin} & High & \priority{100} \\ \hline

 


DBFT & BFT: $3f+1$ & NEO & $<1\text{K}$ & $15-20$sec & \cite{Neo} & Low & \priority{0} \\ \hline


\multirow{4}{*}{DPoS} & \multirow{4}{*}{BFT: $3f+1$} & EOSIO & $1\text{K} - 6\text{K}$ & $<1$sec & \cite{EOSIO} & High & \priority{100} \\
 
& & Bitshares & $100\text{K}$ & $1$sec & \cite{Bitshares} & High & \priority{100} \\
 
& & Tron & $>2\text{K}$ & $3$sec & \cite{Tron} & Med & \priority{50} \\

& & Lisk & $2.5$ & $6$min & \cite{mahankali2019blockchain} & Low & \priority{0} \\

& & Tezos & $\approx 40$ & $\approx 30$min & \cite{Tezos} & Low & \priority{0} \\ \hline

Exonum & BFT: $3f+1$ & Exonum & $\approx 5\text{K}$ & $0.5$sec & \cite{exonum2} & High & \priority{100} \\ \hline
 
IBFT & BFT: $3f+1$ & Quorum & $\approx 600$ & $4.5$sec & \cite{Quorumperformance} & Med & \priority{50} \\ \hline



Kafka & CFT: $2t+1$ & Fabric & $3.5\text{K}$ & $<1$sec & \cite{Androulaki} & High & \priority{75} \\\hline



PoET-SGX & BFT: $\Theta(\frac{\log\log n}{\log n})$ & Sawtooth & $1\text{K} - 2.3\text{K}$ & $<1$sec & \cite{Ampel} & High & \priority{100} \\ \hline
 
\multirow{3}{*}{Raft} & \multirow{3}{*}{CFT: $2t+1$} & Corda & $100-200$& $1$sec & \cite{Ampel} & Low & \priority{0} \\

& & Quorum & $\approx 650$ & $4.5$sec & \cite{Quorumperformance} & Med & \priority{50} \\


& & Fabric & $\approx 7\text{K}$ & $<1$sec & \cite{fabricraft} & High & \priority{75} \\ \hline
 


Ripple & BFT: $4f+1$ & Ripple & $1.5\text{K}$ & $4$sec & \cite{Ripple1} & Med  & \priority{50} \\ \hline


\multirow{2}{*}{Tendermint} & \multirow{2}{*}{BFT: $3f+1$} & Ethermint & $200-800$ & $<1$sec & \cite{Ethermint} & Med  & \priority{50} \\ 
& & Burrow & $>400$ & n/a & \cite{Burrowperformance} & Low & \priority{0} \\  \hline

VBFT & BFT: $3f+1$ & Ontology & $ > 3\text{K}$ & $5-10$sec & \cite{Ontology} & Med  & \priority{50} \\ \hline
 
YAC & BFT: $3f+1$ & Iroha & several $1\text{K}$ & $<3$sec & \cite{YAC}, \cite{iroha1} & Med & \priority{50} \\
\bottomrule
\end{tabular}
\end{table*}

Table \ref{comparisonanalysis}, explains the symbols used below for characterizing the consensus' suitability in IoT networks. With \priority{100} the most prominent BFT protocols having \textit{high performance} (HP) are denoted, whereas \priority{50} is used for BFT protocols with \textit{medium performance} (MP). On the contrary, high performance CFT protocols are appropriate for IoT applications only under the assumption of non-adversarial and honest environment. These assumptions are quite restrictive, focusing on resiliency only, and hence these protocols' suitability \priority{75} is not the maximum possible; for the same reasons, the suitability of CFT protocols with a medium performance is \priority{25}. Finally, fault tolerant and \textit{low performance} (LP) protocols are denoted with \priority{0}\,, implying that they are not suitable to IoT environments. 

Next, a comparative evaluation of 13 consensus protocols and their implementation into blockchain platforms is shown in Table \ref{comparison}; the performance metrics used to determine IoT suitability are fault tolerance, the adoption under permissioned blockchain platforms, transactions' performance, and latency. The transaction's performance is measured in TPS, while the values reported for latency concern either the block confirmation time or the time needed for a transaction to be ordered or notarized, e.g. in \texttt{Fabric}, \texttt{Corda}, etc. The analysis of centralized and PoW based consensus protocols are not included in our survey and in Table \ref{comparison}, as they are either designed for just facilitating development or they are computationally expensive.
The approach used to compute the overall performance in Table \ref{comparison} is that it equals the minimum of the scores achieved by a protocol's throughput and latency.

Although, the above features are evaluated under different setups for different blockchain platforms, the performance of several consensus protocols  is not  mentioned in their white papers. It is expected that, if implemented in a secure, fast, and scalable permissioned blockchain, a high performance consensus protocol should have a promising throughput and low latency. For example, \texttt{Sawtooth} is considered to be a high performance blockchain platform. However, it is not thoroughly tested under different setups in a large scale environment yet. For this reason, such protocols are not currently included in Table \ref{comparison}. The main outcome of the analysis is that even by replacing the monetary concepts with IoT-based criteria, a small number of the proposed consensus protocols meet all the performance and security requirements. With the exception of \texttt{Sawtooth} that is hardware supported by a TEE in its \texttt{PoET-SGX} consensus, the rest of the BFT protocols (\texttt{BFT-SMaRt}, \texttt{DPoS}, and \texttt{Exonum}) achieve quite a high throughput that can reach $100$K TPS and latency as low as $0.5$sec.
 

\section{Conclusions}
\label{sec:conclusion}

In this paper, the potential of using fault-tolerant consensus protocols in IoT networks is investigated. The protocols are analyzed in terms of their performance under different permissioned blockchain platforms and their tolerance against faults, including adversarial ones ---\,a property that is highly desirable in the generally vulnerable IoT ecosystem. The integrated BFT protocols in \texttt{Hyperledger Fabric}, seem to offer a good option for IoT networks when it comes to permissioned blockchains.  

Remarkably, several of the surveyed consensus protocols are under development and not thoroughly tested at the time of writing; for this reason, they have not been included in the comparison. Apart from the performance and fault tolerance that are explored in this paper, several other limitations are faced when blockchain technology is applied in IoT networks. Limitations such as scalability, data privacy, and several other aspects that are part of our going research and will be included in a forthcoming work.


\bibliographystyle{IEEEtran}
\bibliography{Paper}
\end{document}